\documentclass[reqno,twoside,12pt]{amsart}

\hoffset - 1.5cm

\usepackage{amsmath,amssymb}
\usepackage{xcolor}

\newtheorem{df}{Definition}[section]
\newtheorem{thm}[df]{Theorem}
\newtheorem{lem}[df]{Lemma}
\newtheorem{prop}[df]{Proposition}

\newtheorem{alg}[df]{Algorithm}
\newtheorem{exa}[df]{Example}
\title{Wavelet methods in partial differential equations on spheres}
\author{Ilona Iglewska-Nowak}\address{West Pomeranian University of Technology, School of Mathematics, al. Piast\'ow 17, 70--310 Szczecin, Poland, ORCID 0000--0002--1938--8055 } \email{iiglewskanowak@zut.edu.pl}

\author{Piotr Stefaniak}
\address{Faculty of Mathematics and Computer Science
Nicolaus Copernicus University \\
ul. Chopina $12 \slash 18$, PL-87-100 Toru\'{n},
Poland, ORCID 0000--0002--6117--2573} \email{cstefan@mat.umk.pl}

\begin{document}

\begin{abstract}
We propose a method of solving partial differential equations on the $n$-dimen\-sional unit sphere with methods based on the continuous wavelet transform derived from approximate identities.
\end{abstract}


\date{\today}
\maketitle

\section{Introduction}

In the last thirty years, many authors developed wavelet methods for solving differential equations. Already in the 1990s numerical solutions of ODEs \cite{SYYS98,mF99} and PDEs \cite{vP90,sJ92} on the Euclidean space were found. Some further examples of wavelet application to ODEs are presented in \cite{LH14,HMW19}, and to PDEs in \cite{HMV04,MLZJ05,kU09,HK10,cjG14,ZXZ16,BH18}. Contrary to the articles listed so far, the paper \cite{GS04} describes a method for numerical solving of PDEs \emph{on the sphere}. Recent years brought a number of papers in which wavelet methods were involved to solving fractional differential equations \cite{GR14,WMM14,RBAIS20,AAMASD21}. The list is far from being complete, but it is apparent that wavelets are usually being used to develop algorithms for \emph{numerical} solving of differential equations.
No publication is known to us that presents an analytical solution.

Theories of continuous spherical wavelets have been developed in the last decades, simultaneously to theories of wavelets over Euclidean space. Iglewska-Nowak has shown in~\cite[Section~5]{IIN15CWT} that there exist only two essentially different continuous wavelet transforms for spherical signals, namely that based on group theory~\cite{AV,AVn} and that derived from approximate identities~\cite{EBCK09,FGS-book,IIN19CWTforC,IIN15CWT}. In the present paper we show that the latter one can be efficiently applied to solving partial differential equations on the sphere. We give an explicit solution of the Poisson equation, as well as an algorithm for solving the Helmholtz equation.

The present paper seems to be the first attempt to involve wavelet methods to analytical solving of PDEs.
\section{Preliminaries}

\subsection{Functions on the sphere}

A square integrable function~$f$ over the $n$-dimensional unit sphere~$\mathcal S^n\subseteq\mathbb R^{n+1}$, $n\geq 2$, with the rotation-invariant measure~$d\sigma$ normalized such that
$$
\Sigma_n=\int_{\mathcal S^n}d\sigma(x)=\frac{2\pi^{(n+1)/2}}{\Gamma\left((n+1)/2\right)},
$$
can be represented as a Fourier series in terms of the hyperspherical harmonics,
\begin{equation}\label{eq:Fs}
f=\sum_{l=0}^\infty\sum_{k\in\mathcal M_{n-1}(l)}a_l^k(f)\,Y_l^k,
\end{equation}
where $\mathcal M_{n-1}(l)$ denotes the set of sequences $k=(k_0,k_1,\ldots,k_{n-1})$ in $\mathbb N_0^{n-1}\times\mathbb Z$ such that $l\geq k_0\geq k_1\geq\ldots\geq|k_{n-1}|$ and $a_l^k(f)$ are the Fourier coefficients of~$f$. The hyperspherical harmonics of degree~$l$ nad order~$k$ are given by
\begin{equation}\label{eq:Ylk}
Y_l^k(x)=A_l^k\prod_{\tau=1}^{n-1}C_{k_{\tau-1}-k_\tau}^{\frac{n-\tau}{2}+k_\tau}(\cos\vartheta_\tau)\sin^{k_\tau}\!\vartheta_\tau\cdot e^{\pm ik_{n-1}\varphi}
\end{equation}
for some constants $A_l^k$. Here, $(\vartheta_1,\dots,\vartheta_{n-1},\varphi)$ are the hyperspherical coordinates of~$x\in\mathcal S^n$,
\begin{align*}
x_1&=\cos\vartheta_1,\\
x_2&=\sin\vartheta_1\cos\vartheta_2,\\
x_3&=\sin\vartheta_1\sin\vartheta_2\cos\vartheta_3,\\
\dots\\
x_{n-1}&=\sin\vartheta_1\dots\sin\vartheta_{n-2}\cos\vartheta_{n-1},\\
x_n&=\sin\vartheta_1\dots\sin\vartheta_{n-2}\sin\vartheta_{n-1}\cos\varphi,\\
x_{n+1}&=\sin\vartheta_1\dots\sin\vartheta_{n-2}\sin\vartheta_{n-1}\sin\varphi,
\end{align*}
and $\mathcal C_\kappa^K$ are the Gegenbauer polynomials of degree~$\kappa$ and order~$K$.
The set of the hyperspherical harmonics of degree~$l$ is denoted by~$\mathcal H_l$.

The Laplace-Beltrami operator on the sphere $\Delta^{\!\ast}$ is defined by 
\begin{equation*}\index{$\Delta^{\!\ast}$, operator Laplace'a--Bel\-tra\-mi\-ego}
\Delta^{\!\ast}f=\sum\limits_{k=1}^{n-1}\left(\prod\limits_{j=1}^k\sin\vartheta_j\right)^{-2}
 (\sin \vartheta_k)^{k+2-n}\frac{\partial}{\partial\vartheta_k}\left(\sin^{n-k}\vartheta_k\frac{\partial f}{\partial\vartheta_k}\right)
 +\left(\prod\limits_{j=1}^k\sin\vartheta_j\right)^{-2}\frac{\partial^2f}{\partial\varphi^2}.
\end{equation*}
It is known that the hyperspherical harmonics are the eigenfunctions of $\Delta^{\!\ast}$, i.e.,
\begin{equation}\label{eq:HH_eigenvectors}
\Delta^{\!\ast} Y_l^k=-l(l+n-1)Y_l^k,
\end{equation}
see \cite[Chap.~II, Theorem 4.1]{Shimakura}.
The relation of $\Delta^{\!\ast}$ and the Laplace operator $\Delta$ is given by
\begin{equation}\label{eq:laplacians}
\Delta f = R^{-n}\frac{\partial}{\partial R}\left(R^{n}\frac{\partial f}{\partial R}\right) + \frac{1}{R^2}\Delta^{\!\ast} f,
\end{equation}
where
$R\geq 0$ is the radius of $x\in \mathbb{R}^n$ in the hyperspherical coordinates, see \cite[Chap.~II, Proposition 3.3]{Shimakura}.

The Laplace operator is commutative with $SO(n+1)$-rotations $\Upsilon$,
\begin{equation}\label{eq:commutativity}
\Delta\left[f(\Upsilon x)\right]=(\Delta f)(\Upsilon x),
\end{equation}
see \cite[Chap.~IX, \S~2, Subsec.~4]{Vilenkin}. Consequently, from \eqref{eq:laplacians} it follows that the same holds for the Laplace-Beltrami operator, see also \cite[Chap.~II, formula (3.15)]{Shimakura}.

Since $\mathcal{S}^n$ is a manifold without boundary, it follows from the Green's second surface identity that for $f,g$
of class $C^2$ the following holds:
\begin{equation}\label{eq:green}
\int_{\mathcal{S}^n} \Delta^{\!\ast} f(x) \cdot g(x)\, d\sigma(x)=\int_{\mathcal{S}^n}  f(x) \cdot \Delta^{\!\ast}g(x)\, d\sigma(x).
\end{equation}

The scalar product in $\mathcal L^2(\mathcal S^n)$ is antilinear in the first variable,
$$
\left<f,g\right>=\frac{1}{\Sigma_n}\int_{\mathcal{S}^n}\overline{f(x)}\,g(x)\,d\sigma(x).
$$
Since~$\Delta^{\!\ast}$ is a linear operator, one has
$$
\overline{\Delta^{\!\ast} f}=\Delta^{\!\ast}\overline{f}
$$
and~\eqref{eq:green} can be also written as
\begin{equation}\label{eq:Green_sp}
\left<\Delta^{\!\ast}f,g\right>=\left<f,\Delta^{\!\ast}g\right>.
\end{equation}

Zonal (rotation-invariant) functions are those depending only on the first hyperspherical coordinate $\vartheta=\vartheta_1$. Unless it leads to misunderstandings, we identify them with functions of~$\vartheta$ or $t=\cos\vartheta$.
A zonal $\mathcal L^2$-function~$f$ has the following Gegenbauer expansion
\begin{equation}\label{eq:Gegenbauer_expansion}
f(t)=\sum_{l=0}^\infty\widehat f(l)\,C_l^\lambda(t),\qquad t=\cos\vartheta,
\end{equation}
where $\widehat f(l)$ are the Gegenbauer coefficients of~$f$ and $\lambda$ is related to the space dimension by
$$
\lambda=\frac{n-1}{2}.
$$
Consequently, for a zonal $\mathcal L^2$-function $f$ one has
$$
\widehat f(l)=A_l^0\cdot a_l^0(f),
$$
compare~\eqref{eq:Fs}, \eqref{eq:Ylk}, and~\eqref{eq:Gegenbauer_expansion}.

For $f,g\in\mathcal L^1(\mathcal S^n)$, $g$ zonal, their convolution $f\ast g$ is defined by~\cite[Definition~2.1.1]{DX13}
$$
(f\ast g)(x)=\frac{1}{\Sigma_n}\int_{\mathcal S^n}f(y)\,g(x\cdot y)\,d\sigma(y).
$$
With this notation one has
$$
f\in\mathcal H_l\Longrightarrow f=\frac{\lambda+l}{\lambda}\left(f\ast C_l^\lambda\right)
$$
(Funck-Hecke formula), i.e., the function $\frac{\lambda+l}{\lambda}\, C_l^\lambda$ is the reproducing kernel for~$\mathcal H_l$.

\subsection{Frames in reproducing kernel Hilbert spaces}

One of the wavelet definitions considered in the present paper requires the notion of a continuous frame. Here we present the one given in~\cite{gK94}.

\begin{df}Let~$\mathcal H$ be a Hilbert space and let~$M$ be a measure space with measure~$\mu$. A generalized frame in~$\mathcal H$ indexed by~$M$ is a family of vectors $\mathcal \{h_m\in\mathcal H:\,m\in M\}$ such that
\begin{itemize}
\item[(a)] For every $f\in\mathcal H$, the function $\tilde f:\,M\to\mathbb C$ defined by
$$
\tilde f(m)=\left<h_m,f\right>_{\mathcal H}
$$
is measurable.
\item[(b)] There exists a pair of constants $0<A<B<\infty$ such that for every $f\in\mathcal H$,
\begin{equation}\label{eq:frame_condition}
A\|f\|_\mathcal H^2\leq\|\tilde f\|_{\mathcal L^2(M)}^2\leq B\|f\|_\mathcal H^2.
\end{equation}
\end{itemize}
\end{df}

The mapping
\begin{align*}
T\colon\,\mathcal H&\to\mathcal L^2(M)\\
f&\mapsto\tilde f
\end{align*} is a linear operator, which is called the frame operator. By the frame condition~\eqref{eq:frame_condition} it is bounded and invertible.

\begin{bfseries}Remark.\end{bfseries} In the theory of discrete frames which are much more popular than continuous ones, see, e.g., \cite{oC03,CJ99,kG01}, operator~$T$ is called \emph{analysis} operator or \emph{pre-frame operator}, whereas \emph{frame operator} means $T^\ast T$.

Theorem~4.4 in~\cite{gK94} states the following.

\begin{thm}
The synthesizing operator $S=(T^\ast T)^{-1}T^\ast:\,\mathcal L^2(M)\to\mathcal H$ is given by
$$
Sg=\int_Mh^m g(m)\,d\nu(m),
$$
where
$$
h^m=(T^\ast T)^{-1}h_m
$$
is the reciprocal frame of~$\mathcal H_M$. In particular, $f\in\mathcal H$ can be reconstructed from $\tilde f\in\mathcal F$ by
$$
f=S\tilde f=\int_Mh^m\tilde f(m)\,d\nu(m).
$$
\end{thm}

On the other hand, signal reconstruction from its frame coefficients can be done iteratively by the so-called frame algorithm, compare~\cite[Section~5.1]{kG01}.

\begin{alg}
Given a relaxation parameter $0<\varrho<\frac{2}{B}$, set $\delta=\max\{|1-\varrho A|,|1-\varrho B|\}<1$. Let $f_0=0$ and define recursively
$$
f_{k+1}=f_k+\lambda(T^\ast T)(f-f_k).
$$
Then, $\lim_{k\to\infty}f_k=f$ with a geometric rate of convergence,
$$
\|f-f_k\|\leq\delta^k\|f\|.
$$
Note that
$$
f_1=\varrho(T^\ast T)f=\varrho\int_M h_m\left<h_m,f\right>\,d\mu(m)
$$
contains the frame coefficients as input. This suffices to compute the further approximations~$f_k$ and to reconstruct~$f$.
\end{alg}

\section{Two wavelet transforms}\label{sec:bwt}

A wavelet transform based on approximate identities can be performed in a twofold way. In the first case, the constraints on a function to be a wavelet are quite restrictive, but the inverse transform is given directly by an integral. This is the more popular version of the wavelet transform, developed starting from the 1990s by Freeden \emph{et al.} \cite{FGS-book,FW-C,FW} and Bernstein \emph{et al.} \cite{sB09,EBCK09}, as well as by Iglewska-Nowak in the recent years \cite{IIN19CWTforC,IIN15CWT,IIN18DW,IIN17FDW}.

The following definition comes from \cite{IIN19CWTforC}, it is an improved version of the one used in \cite{IIN15CWT}, adapted to the case of zonal wavelets and with the most popular weight function $\alpha(\rho)=\frac1\rho$.

\begin{df}\label{def:wavelet_1} A family $\{\Psi_\rho\}_{\rho\in\mathbb R_+}\subseteq\mathcal L^2(\mathcal S^n)$ of rotation-invariant functions is called an admissible spherical wavelet  if it satisfies the following condition:
\begin{equation}\label{eq:admbwv}
\int_0^\infty\left|\widehat{\Psi_\rho}(l)\right|^2\,\frac{d\rho}{\rho}=\left(\frac{\lambda+l}{\lambda}\right)^2\qquad\text{for }l\in\mathbb{N}_0.
\end{equation}
\end{df}

The wavelet transform is defined by
\begin{equation}\label{eq:bwt}
\mathcal W_\Psi f(\rho,y)=\frac{1}{\Sigma_n}\int_{\mathcal{S}^n}\overline{\Psi_\rho(y\cdot x)}\cdot f(x)\,d\sigma(x)=\left<\tau_y\Psi_\rho,f\right>,
\end{equation}
where~$\tau_y$ denotes the translation operator of zonal functions,
$$
\tau_yf(x)=f(y\cdot x).
$$
Note that this operator can be represented by
$$
\tau_yf(x)=f(A_y\cdot x),
$$
where $A_y$ is an $SO(n+1)$-matrix, corresponding to the translation $\tau_y$.

The wavelet transform is invertible by an integral. Theorem~3.2 in \cite{IIN19CWTforC} reduced to the case of zonal functions states the following.

\begin{thm}\label{thm:inversion_1} If $\{\Psi_\rho\}_{\rho\in\mathbb R_+}$ is an admissible wavelet, then the wavelet transform is invertible by
\begin{equation}\label{eq:bwt_synthesis}
f(x)=\frac{1}{4\pi}\int_0^\infty\!\!\int_{\mathcal S^n}\Psi_\rho(x\cdot y)\cdot\mathcal W_\Psi\,f(\rho,y)\,d\sigma(y)\,\frac{d\rho}{\rho}
\end{equation}
with the limit in $\mathcal L^2$-sense.
\end{thm}

Note that the wavelet transform definition from~\cite[formula~(4)]{IIN19CWTforC} should be corrected by the factor~$\frac{1}{\Sigma_n}$ such that it coincides with the one used in the present paper. The article~\cite{IIN19CWTforC} is based on~\cite{IIN15CWT}, where the wavelet transform is defined in the same way as in formula~\eqref{eq:bwt}. In the proof of \cite[Theorem~3.2]{IIN19CWTforC} the reproducing property of the kernels $\frac{\lambda+l}{\lambda}\mathcal C_l^\lambda$ is used (falsely) without the factor $\frac{1}{\Sigma_n}$.

Examples of wavelets satisfying Definition~\ref{def:wavelet_1} are first of all the Gauss-Weierstrass wavelet given by
$$
\widehat{\Psi_\rho^G}(l)=\sqrt{2l(l+2\lambda)\rho}\,e^{-l(l+2\lambda)\rho}\cdot\frac{\lambda+l}{\lambda},\quad l\in\mathbb N_0,
$$
the Abel-Poisson wavelet,
$$
\widehat{\Psi_\rho^A}(l)=\sqrt{2l\rho}\,e^{-l\rho}\cdot\frac{\lambda+l}{\lambda},\quad l\in\mathbb N_0,
$$
and the Poisson wavelets of order $d\in\mathbb N$,
$$
\widehat{\Psi_\rho^d}(l)=\frac{2^d}{\sqrt{\Gamma(2d)}}\cdot(l\rho)^d\,e^{-l\rho}\cdot\frac{\lambda+l}{\lambda},\quad l\in\mathbb N_0.
$$
For more details on the latter family see~\cite{IIN15PW}.

Another definition of a spherical wavelet transform based on approximate identities has been introduced in~\cite{IIN17FDW} in order to obtain a wider class of functions that can be used as wavelets. The constraints on a wavelets family are weaker as in Definition~\ref{def:wavelet_1} but the price to be paid is the lack of a direct inverse wavelet transform. The analysed signal can be reconstructed by frame methods.

\begin{df}\label{def:wavelet_2} The family $\{\Psi_\rho\}_{\rho\in\mathbb R_+}\subseteq\mathcal L^2(\mathcal S^n)$ is called a wavelet (family) of order~$m$ if $\widehat{\Psi_\rho}(l)=0$ for $l=0,1,\ldots,m$ and it satisfies
\begin{equation}\label{eq:cond_energy_conservation}
A\cdot\left(\frac{\lambda+l}{\lambda}\right)^2\leq\int_0^\infty|\widehat{\Psi_\rho}(l)|^2\,\frac{d\rho}{\rho}\leq B\cdot\left(\frac{\lambda+l}{\lambda}\right)^2
\end{equation}
for some positive constants~$A$ and~$B$ independent of $l\in\mathbb{N}_0$ and $l>m$. The constants $A$ and $B$ are called wavelet family bounds of the wavelet family $\{\Psi_\rho\}$.
\end{df}

The wavelet transform is performed according to~\eqref{eq:bwt} and it follows that the wavelet coefficients $\{\mathcal W_\Psi f(\rho,y)\}$ constitute a frame (see~\cite[Theorem~2.5]{IIN17FDW}).

\begin{thm}\label{thm:WT_frame_property} Let $\{\Psi_\rho\}$ be a wavelet family of order~$m$. Then for any $f\in\mathcal L^2(\mathcal S^n)$ with $m$ vanishing moments (i.e., such that $a_l^k(f)=0$ for $l=0,1,\dots,m$ and $k\in\mathcal M_{n-1}(l)$) we have
$$
A\|f\|^2\leq\|\mathcal W_\Psi f\|^2\leq B\|f\|^2,
$$
i.e., the set $\{\tau_y\Psi_\rho,\,\rho\in(0,\infty),y\in\mathcal S^n\}$ is a frame for $\mathcal L^2(\mathcal S^n)\setminus\bigcup_{l=0}^m\mathcal H_l$.
\end{thm}

This means that in this case the wavelet transform is invertible, but with use of frame techniques.

\section{Differential equations}

\subsection{The Poisson equation $\Delta^{\!\ast} u=f$}

Consider the Poisson equation
\begin{equation}\label{eq:Poisson_eq}
\Delta^{\!\ast} u=f,
\end{equation}
where both functions~$u$ and~$f$ are given as the Fourier series
$$
u=\sum_{l=0}^\infty\sum_{k\in\mathcal M_{n-1}(l)}a_l^k(u)\,Y_l^k,\qquad f=\sum_{l=0}^\infty\sum_{k\in\mathcal M_{n-1}(l)}a_l^k(f)\,Y_l^k.
$$
Since the hyperspherical harmonics~$Y_l^k$ are linearly independent, one obtains by~\eqref{eq:HH_eigenvectors} the following relation
$$
-l(l+n-1)\,a_l^k(u)=a_l^k(f).
$$
If $a_0^0(f)=0$,
\begin{equation}\label{eq:Laplace_eq_solution_series}
u=\sum_{l=1}^\infty\sum_{k\in\mathcal M_{n-1}(l)}\frac{-a_l^k(f)}{l(l+n-1)}\cdot Y_l^k
\end{equation}
is a solution to the equation~\eqref{eq:Poisson_eq}. This statement is the content of~\cite[Theorem~2.1]{BZ09}, however, convergence of the series is not proved. Wavelet methods yield  the same formula for the solution of~\eqref{eq:Poisson_eq} and, additionally, the proof of convergence.

\begin{thm}Let $f$ be an $\mathcal L^1(\mathcal S^n)$-function such that $a_0^0(f)=0$. Then, the function
$$
u=f\ast\mathcal K
$$
with
\begin{equation}\label{eq:kernel_series}
\mathcal K=\sum_{l=1}^\infty\frac{-1}{l(n+l-1)}\cdot\frac{\lambda+l}{\lambda}\,\mathcal C_l^\lambda
\end{equation}
is a solution to the equation
$$
\Delta^{\!\ast}u=f.
$$
\end{thm}

\begin{bfseries}Proof.\end{bfseries} Consider the family of functions
\begin{equation}\label{eq:psirhod}
\psi_\rho^d
=\sum_{l=0}^\infty\underbrace{\frac{\rho^d l^{d+1}}{l+n-1}\cdot e^{-\rho l}\cdot\frac{\lambda+l}{\lambda}}_{\widehat{\psi_\rho^d}(l)}\,\mathcal C_l^\lambda
=\frac{\sqrt{\Gamma(2d)}}{2^d}\cdot\sum_{l=0}^\infty\frac{l}{l+n-1}\cdot \widehat{\Psi_\rho^d}(l)\,\mathcal C_l^\lambda  .
\end{equation}
(Recall that $\Psi_\rho^d$ is -- the Poisson wavelet of order~$d$). Its coefficients satisfy
\begin{align*}
\int_0^\infty|\widehat{\psi_\rho^d}(l)|^2\frac{d\rho}{\rho}
&=\frac{\Gamma(2d)\cdot l^2}{4^d\cdot(l+n-1)^2}\cdot\left(\frac{\lambda+l}{\lambda}\right)^2,
\end{align*}
i.e., $\{\psi_\rho^d\}$ is a wavelet with respect to Definition~\ref{def:wavelet_2}. Wavelet family bounds can be chosen to be equal to
$$
A:=\frac{\Gamma(2d)}{4^d}\cdot\inf_{l\in\mathbb N}\frac{l^2}{(l+n-1)^2}=\frac{\Gamma(2d)}{4^d\cdot n^2},\qquad
   B:=\frac{\Gamma(2d)}{4^d}\cdot\sup_{l\in\mathbb N}\frac{l^2}{(l+n-1)^2}=\frac{\Gamma(2d)}{4^d}.
$$

The wavelet transform of~$f$ is given by
$$
\mathcal W_{\psi^d} f(\rho,y)=\left<\tau_y\psi_\rho^d,f\right>=\left<\tau_y\psi_\rho^d,\Delta^{\!\ast} u\right>.
$$
From \eqref{eq:Green_sp} and the differentiability of~\eqref{eq:psirhod} it follows that
\begin{equation}\label{eq:WT_f_u}
\mathcal W_{\psi^d} f(\rho,y)=\left<\Delta^{\!\ast}(\tau_y\psi_\rho^d),u\right>.
\end{equation}
Set~$\theta_\rho^d:=\rho^2\Delta^{\!\ast}\psi_\rho^d$. According to~\eqref{eq:HH_eigenvectors} and \eqref{eq:Gegenbauer_expansion}, the coefficients of this family satisfy
$$
\widehat{\theta_\rho^d}(l)=-\rho^2l(l+n-1)\,\widehat{\psi_\rho^d}(l)=-(\rho l)^{d+2}\,e^{-\rho l}\cdot\frac{\lambda+l}{\lambda},
$$
i.e., $\{\theta_\rho^d\}$ is the Poisson wavelet of order~$d+2$ (up to a constant). By~\eqref{eq:WT_f_u} and the commutativity of the Laplace-Beltrami operator with rotations~\eqref{eq:commutativity},
\begin{equation}\label{eq:RelationOfTransforms}
\rho^2\cdot\mathcal W_{\psi^d}f=\left<\tau_y(\rho^2\Delta^{\!\ast}\psi_\rho^d),u\right>=\mathcal W_{\theta^d}u.
\end{equation}
Since $\{-\frac{2^{d+2}}{\sqrt{\Gamma(2d+4)}}\cdot\theta_\rho^d\}$ is a wavelet with respect to Definition~\ref{def:wavelet_1}, direct inversion is possible, i.e.,
$$
u(x)=\frac{4^{d+2}}{\Sigma_n\cdot\Gamma(2d+4)}
   \cdot\int_0^\infty\!\int_{\mathcal S^n}\theta_\rho^d(x\cdot y)\cdot\mathcal W_{\theta^d}u(\rho,y)\,d\sigma(y)\,\frac{d\rho}{\rho}.
$$
Hence, by \eqref{eq:RelationOfTransforms},
$$
u(x)=\frac{4^{d+2}}{\Sigma_n\cdot\Gamma(2d+4)}
   \cdot\int_0^\infty\!\int_{\mathcal S^n}\theta_\rho^d(x\cdot y)\cdot\rho^2\,\mathcal W_{\psi^d}f(\rho,y)\,d\sigma(y)\,\frac{d\rho}{\rho}.
$$
Substitute the right-hand-side of~\eqref{eq:bwt} for~$\mathcal W_{\psi^d}f$ to obtain
$$
u(x)=\frac{4^{d+2}}{\Sigma_n^2\cdot\Gamma(2d+4)}\cdot\int_0^\infty\!\int_{\mathcal S^n}\theta_\rho^d(x\cdot y)
   \cdot\rho^2\,\int_{\mathcal S^n}\overline{\psi^d(y\cdot z)}\, f(z)\,d\sigma(z)\,d\sigma(y)\,\frac{d\rho}{\rho}.
$$
The triple integral is convergent in $\mathcal L^2$-sense (with $\int_0^\infty$ understood as $\lim_{R\to0}\int_R^{1/R}$), compare the proof of~\cite[Theorem~3.2]{IIN19CWTforC}. Thus, the order of integration may be changed,
\begin{equation}\label{eq:u_as_triple_integral}
u(x)=\frac{4^{d+2}}{\Sigma_n^2\cdot\Gamma(2d+4)}\cdot\int_{\mathcal S^n}f(z)\int_0^\infty
   \int_{\mathcal S^n}\rho\cdot\theta_\rho^d(x\cdot y)\cdot\overline{\psi^d(y\cdot z)}\,d\sigma(y)\,d\rho\,d\sigma(z).
\end{equation}
Now, \cite[formula~(15)]{IIN15CWT} written for two zonal functions~$f$, $h$,
$$
f\ast h=\sum_{l=0}^\infty\frac{n-1}{n+2l-1}\,\widehat f(l)\,\widehat h(l)\,\mathcal C_l^\lambda,
$$
yields
\begin{align}
&\frac{\rho}{\Sigma_n}\cdot\int_{\mathcal S^n}\theta_\rho^d(x\cdot y)\cdot\overline{\psi^d(y\cdot z)}\,d\sigma(y)\label{eq:convolution_theta_psi}\\
&=\rho\cdot\sum_{l=0}^\infty\frac{n-1}{n+2l-1}\cdot\underbrace{\left[-(\rho l)^{d+2}\cdot e^{-\rho l}\cdot\frac{n+2l-1}{n-1}\right]}_{=\widehat{\theta_\rho^d}(l)}
   \cdot\underbrace{\frac{\rho^dl^{d+1}}{n+l-1}\cdot e^{-\rho l}\cdot\frac{n+2l-1}{n-1}}_{=\widehat{\psi_\rho^d}(l)}\cdot\,\mathcal C_l^\lambda(x\cdot z)\notag\\
&=-\sum_{l=0}^\infty\frac{(\rho l)^{2d+3}}{n+l-1}\cdot e^{-2\rho l}\cdot\frac{\lambda+l}{\lambda}\,\mathcal C_l^\lambda(x\cdot z).\label{eq:convolution_theta_psi}
\end{align}
Further,
\begin{equation}\label{eq:kernel_coefficients}
\int_0^\infty(\rho l)^{2d+3}\cdot e^{-2\rho l}\,d\rho=\frac{\Gamma(2d+4)}{4^{d+1}\cdot l}.
\end{equation}
Substitute~\eqref{eq:convolution_theta_psi} and~\eqref{eq:kernel_coefficients} into~\eqref{eq:u_as_triple_integral} to obtain
$$
u(x)=\frac{1}{\Sigma_n}\cdot\int_{\mathcal S^n}\mathcal K(x\cdot z)\cdot f(z)\,d\sigma(z)=f\ast\mathcal K(x)
$$
with
\begin{align}
\mathcal K&=-\frac{4^{d+2}}{\Gamma(2d+4)}\cdot\int_0^\infty\sum_{l=0}^\infty\frac{(\rho l)^{2d+3}}{n+l-1}
   \cdot e^{-2\rho l}\cdot\frac{\lambda+l}{\lambda}\,\mathcal C_l^\lambda\,d\rho\label{eq:K_as_integral}\\
&=\sum_{l=1}^\infty\frac{-1}{l(n+l-1)}\cdot\frac{\lambda+l}{\lambda}\,\mathcal C_l^\lambda.\label{eq:K_series}
\end{align}
\hfill$\Box$

\begin{bfseries}Remark.\end{bfseries} Since $\frac{\lambda+l}{\lambda}\,\mathcal C_l^\lambda$ is the reproducing kernel for~$\mathcal H_l$, for~$f$ given as Fourier series this yields~\eqref{eq:Laplace_eq_solution_series}. Convergence of the series representing both~$\mathcal K$ and~$f$ follows from the arguments given in the proof of~\cite[Theorem~3.2]{IIN19CWTforC}. Note that~$\mathcal K$ is in principle a convolution of two Poisson wavelets, integrated with respect to the scale parameter. Thus, it is twice continuously differentiable. This ensures that $u\in\mathcal C^2(\mathcal S^n)$.\\

\subsubsection{Direct representations of the kernel~$\mathcal K$}\label{subs:direct_K}

Now, in order to obtain a direct expression for~$\mathcal K$, note that~\eqref{eq:K_series} can be obtained from the Poisson kernel \cite[formulae~(4) and~(5)]{IIN15PW}
\begin{equation}\label{eq:Poisson_kernel}
p_r(\cos\vartheta)=\frac{1}{\Sigma_n}\cdot \frac{1-r^2}{(1-2r\cos\vartheta+r^2)^{(n+1)/2}}
   =\frac{1}{\Sigma_n}\cdot\sum_{l=0}^\infty r^l\cdot\frac{\lambda+l}{\lambda}\,\mathcal C_l^\lambda(\cos\vartheta).
\end{equation}

\begin{prop}\label{prop:direct_K} The kernel~$\mathcal K$, given by~\eqref{eq:kernel_series}, is equal to
\begin{equation}\label{eq:kernel_integral}
\mathcal K(\cos\vartheta)
   =-\lim_{\epsilon\to0}\int_0^{1-\epsilon}R^{n-2}\cdot\int_0^R\frac{\Sigma_n\cdot p_r(\cos\vartheta)-1}{r}\,dr\,dR.
\end{equation}
\end{prop}

{\bf Remark.} The upper bound in the outer integral cannot be chosen to be equal to~$1$, because $p_r$ is defined only for $r\in[0,1)$. However, the integrand $\frac{\Sigma_n\cdot p_r(\cos\vartheta)-1}{r}$ can be continuously extended to $r=1$, such that actually, the second integration is performed over the interval $[0,1]$.

\begin{bfseries}Proof. \end{bfseries} Since the Gegenbauer polynomials~$\mathcal C_l^\lambda$ over the interval $[-1,1]$ are bounded by
\begin{equation}\label{eq:Cllambda_bound}
\left|\mathcal C_l^\lambda(\cos\vartheta)\right|\leq(n+l-2)^{n-2}
\end{equation}
uniformly in $\vartheta\in[0,\pi]$ (compare \cite[Theorem~7.33.1]{Sz75}), the series~\eqref{eq:Poisson_kernel} is absolutely convergent for $r\in[0,1)$. Note that for $n\geq2$ and $l\geq1$
$$
\frac{1}{l(n+l-1)}=\int_0^1R^{n-2}\cdot\int_0^R r^{l-1}\,dr\,dR.
$$
Substitute this expression to~\eqref{eq:kernel_series} to obtain~\eqref{eq:kernel_integral}, taking into account that $\mathcal C_0^\lambda(\cos0)=1$.\hfill$\Box$

\begin{exa} For $n=2$ it follows from~\eqref{eq:Poisson_kernel}
$$
\beta_r(t):=\sum_{l=1}^\infty r^{l-1}\cdot(2l+1)\,\mathcal C_l^1(t)=\frac{1-r^2}{r(1-2tr+r^2)^{3/2}}-\frac{1}{r}.
$$
Further,
$$
\gamma_r(t):=\int\beta_r(t)\,dr=\frac{2}{\sqrt{1-2tr+r^2}}-\ln\left(1-rt+\sqrt{1-2tr+r^2}\right)+C
$$
and
$$
\zeta_R(t):=\int[\gamma_R(t)-\gamma_0(t)]\,dR=\ln\left(R-t+\sqrt{1-2tR+R^2}\right)-R\left(1+\ln\frac{1-tR+\sqrt{1-2tR+R^2}}{2}\right).
$$
Consequently,
$$
\mathcal K^{(2)}(t)=\zeta_0^{(2)}(t)-\zeta_1^{(2)}(t)=1+\ln\frac{1-t}{2}.
$$
For $t=\cos\vartheta$ it can be expressed as
$$
\mathcal K^{(2)}(\cos\vartheta)=1+\ln\left(\left(\sin\frac{\vartheta}{2}\right)^2\right).
$$
\end{exa}

Table~\ref{tab:kernel_K} gives the expressions for~$\mathcal K$ for $n=2,3,4,5,6,7,8,9,10$.

\begin{table}\label{tab:kernel_K}\caption{Kernel~$\mathcal K$ for different values of~$n$, $t=\cos\vartheta$}\vspace{0.5em}\centering\begin{tabular}{|l|l|}
\hline
$n=2$&$1+2\ln\left(\frac{1-t}{2}\right)$\\[0.5em]
$n=3$&$\frac{-(\pi-\vartheta)\cdot t}{2\sqrt{1-t^2}}+\frac{1}{4}$\\[0.5em]
$n=4$&$\frac{4-7t}{9(1-t)}+\frac{1}{3}\ln\left(\frac{1-t}{2}\right)$\\[0.5em]
$n=5$&$\frac{(\pi-\vartheta)\cdot t(3-2t^2)}{2(1-t^2)^{3/2}}-\frac{3-5t^2}{16(1-t^2)}$\\[0.5em]
$n=6$&$\frac{23-71t+43t^2}{75(1-t)^2}+\frac{1}{5}\ln\left(\frac{1-t}{2}\right)$\\[0.5em]
$n=7$&$\frac{(\pi-\vartheta)\cdot(-15+20t^2-8t^4)}{48(1-t^2)^{5/2}}+\frac{22-71t^2+40t^4}{144(1-t^2)^2}$\\[0.5em]
$n=8$&$\frac{176-759t+906t^2-337t^3}{735(1-t)^3}+\frac{1}{7}\ln\left(\frac{1-t}{2}\right)$\\[0.5em]
$n=9$&$\frac{(\pi-\vartheta)\cdot(-35+70t^2-56t^4+16t^6)}{128(1-t^2)^{7/2}}+\frac{50-237t^2+266t^4-94t^6}{384(1-t^2)^3}$\\[0.5em]
$n=10$&$\frac{563-3089t+5466t^2-4049t^3+1091t^4}{2835(1-t)^4}+\frac{1}{9}\ln\left(\frac{1-t}{2}\right)$\\\hline
\end{tabular}\end{table}

\subsection{The Helmholtz equation $\Delta^{\!\ast} u+au=f$, $a\in\mathbb C$}

\subsubsection{Solution with wavelet methods}

Let $\{\Psi_\rho^d\}$ be the Poisson wavelet family of order~$d$, i.e.,
$$
\Psi_\rho^d=\sum_{l=0}^\infty(\rho l)^d\cdot e^{-\rho l}\cdot\frac{\lambda+l}{\lambda}\,\mathcal C_l^\lambda.
$$
The wavelet transform of~$f$ is equal to
$$
\mathcal W_{\Psi^d} f(\rho,y)=\left<\tau_y\Psi_\rho^d,f\right>=\left<\tau_y\Psi_\rho^d,\Delta^{\!\ast} u\right>+\left<\tau_y\Psi_\rho^d,au\right>,
$$
and it can be written as
$$
\mathcal W_{\Psi^d} f(\rho,y)=\left<\tau_y[(\Delta^{\!\ast}\!+\bar a)\Psi_\rho^d],u\right>
$$
(by the commutativity of the Laplace-Beltrami operator with rotations~\eqref{eq:commutativity}). The Gegenbauer coefficients of the family $\Theta_\rho^d:=\rho^2(\Delta^{\!\ast}\!+\bar a)\Psi_\rho^d$ are equal to
\begin{align*}
\widehat{\Theta_\rho^d}(l)&=\rho^2\left[-l(l+n-1)+\bar a\right]\widehat{\Psi_\rho^d}(l)\\
&=-\widehat{\Psi_\rho^{d+2}}(l)-\rho\cdot(n-1)\,\widehat{\Psi_\rho^{d+1}}(l)+\rho^2\cdot\bar a\,\widehat{\Psi_\rho^d}(l)
\end{align*}
and they satisfy
$$
\int_0^\infty\left|\widehat{\Theta_\rho^d}(l)\right|^2\,\frac{d\rho}{\rho}=\frac{\Gamma(2d+4)}{2^{2(d+2)}}\cdot\frac{[l(l+n-1)-\bar a]^2}{l^4}
   \cdot\left(\frac{\lambda+l}{\lambda}\right)^2.
$$
If $a\ne l(l+n-1)$ for all $l\in\mathbb N$, the family $\{\Theta_\rho^d\}$ is a wavelet according to Definition~\ref{def:wavelet_2}. In this case,
$$
\rho^2\cdot\mathcal W_{\Psi^d}f(\rho,y)=\left<\tau_y\Theta_\rho^d,u\right>=\mathcal W_{\Theta^d}u(\rho,y)
$$
and the function~$u$ can be recovered by frame methods.

\subsubsection{The generalized Green function for the Helmholtz equation}

The Helmholtz equation is a subject of research of Szmytkowski~\cite{rSz06,rSz07}. The author studies the case when $a=L(L+n-1)$, $L\in\mathbb Z$. It turns out that the solution is given by
$$
u=\int_{\mathcal S^n}G_L(x\cdot y)\cdot f(y)\,d\sigma(y).
$$

One has the generalized Green function
$$
G_L(x\cdot y)=-\sum_{\genfrac{}{}{0pt}{2}{l=0}{l\ne L}}^\infty\sum_{k\in\mathcal M_{n-1}(l)}\frac{Y_l^k(x)\cdot\overline{Y_l^k(y)}}{l(l+n-1)-L(L+n-1)}.
$$
In the papers \cite{rSz06,rSz07} closed forms of these functions are given. With the method developed in Subsection~\ref{subs:direct_K}, we are able to derive an algorithm for finding direct expressions for~$G_L$. It is different from the one presented by Szmytkowski. For the set of indices that we have tested, the expressions for~$G_L$ coincide with those in~\cite{rSz06,rSz07}.

\begin{thm}
Let $n\geq2$ be fixed. Then,
\begin{equation}\label{eq:GL}\begin{split}
G_L(\cos\vartheta)&=-\lim_{\epsilon\to0}\int_0^{1-\epsilon}R^{n+2k}\int_0^Rr^{n+k-2}
   \cdot\left(\Sigma_n\cdot p_r(\cos\vartheta)-\sum_{l=0}^Lr^l\cdot\frac{\lambda+l}{\lambda}\,\mathcal C_l^\lambda(\cos\vartheta)\right)\,dr\,dR\\
&-\sum_{l=0}^{L-1}\frac{1}{(l-L)(l+n+L-1)}\cdot\frac{\lambda+l}{\lambda}\,\mathcal C_l^\lambda(\cos\vartheta).
\end{split}\end{equation}
\end{thm}

\begin{bfseries}Proof.\end{bfseries} Note that
$$
\sum_{k\in\mathcal M_{n-1}(l)}Y_l^k(x)\cdot\overline{Y_l^k(y)}=\frac{\lambda+l}{\lambda}\,\mathcal C_l^\lambda(x\cdot y)
$$
(\cite[Theorem 1.2.6]{DX13} together with \cite[Lemma 1.2.3]{DX13}) and
$$
l(l+n-1)-L(L+n-1)=(l-L)(l+n+L-1).
$$
Further,
$$
\Sigma_n\cdot p_r(\cos\vartheta)-\sum_{l=0}^Lr^l\cdot\frac{\lambda+l}{\lambda}\,\mathcal C_l^\lambda(\cos\vartheta)
   =\sum_{l=L+1}^\infty r^l\cdot\frac{\lambda+l}{\lambda}\,\mathcal C_l^\lambda(\cos\vartheta).
$$
The statement~\eqref{eq:GL} follows from
$$
\frac{1}{(l-L)(l+n+L-1)}=\int_0^1R^{-(n+2k)}\cdot\int_0^R r^{n+L-2}\cdot r^l\,dr\,dR
$$
with the same arguments as in the proof of Proposition~\ref{prop:direct_K}.\hfill$\Box$

Table~\ref{tab:GL} gives the expressions for~$G_L(\cos\vartheta)$ for several values of parameters $n$ and $L$. They coincide with those derived in~\cite{rSz06} and~\cite{rSz07}.

\begin{table}\label{tab:GL}\caption{The Green function~$G_L$ for different values of~$n$, $L$; $t=\cos\vartheta$}\vspace{0.5em}\centering\begin{tabular}{|r|l|}
\hline
$n=2$, $L=1$ &$1+\frac{4}{3}t+t\cdot\ln\left(\frac{1-t}{2}\right)$\\[0.5em]
$n=2$, $L=2$ &$\frac{-7+30t+41t^2}{20}-\frac{1-3t^2}{2}\cdot\ln\left(\frac{1-t}{2}\right)$\\[0.5em]
$n=2$, $L=3$ &$\frac{-56-123t+210t^2+289t^3}{84}-\frac{t(3-5t^2)}{2}\cdot\ln\left(\frac{1-t}{2}\right)$\\[0.5em]
$n=2$, $L=4$ &$\frac{75-660t-1182t^2+1260t^3+1739t^4}{288}+\frac{3-30t^2+35t^4}{8}\cdot\ln\left(\frac{1-t}{2}\right)$\\[0.5em]
$n=3$, $L=1$ &$\frac{(\pi-\vartheta)\cdot(1-2t^2)}{2\sqrt{1-t^2}}+\frac{t}{4}$\\[0.5em]
$n=3$, $L=2$ &$\frac{(\pi-\vartheta)\cdot t(3-4t^2)}{2\sqrt{1-t^2}}-\frac{1-4t^2}{12}$\\[0.5em]
$n=3$, $L=3$ &$\frac{(\pi-\vartheta)\cdot(-1+8t^2-8t^4)}{2\sqrt{1-t^2}}-\frac{t(1-2t^2)}{4}$\\[0.5em]
$n=3$, $L=4$ &$\frac{(\pi-\vartheta)\cdot t(-5+20t^2-16t^4)}{2\sqrt{1-t^2}}+\frac{1-12t^2+16t^4}{20}$\\[0.5em]
$n=4$, $L=1$ &$\frac{10+13t-28t^2}{15(1-t)}+t\cdot\ln\left(\frac{1-t}{2}\right)$\\[0.5em]
$n=4$, $L=2$ &$\frac{-41+223t+149t^2-359t^3}{84(1-t)}-\frac{1-5t^2}{2}\cdot\ln\left(\frac{1-t}{2}\right)$\\[0.5em]
$n=4$, $L=3$&$\frac{-96-213t+903t^2+397t^3-1027t^4}{108(1-t)}-\frac{5t(3-7t^2)}{6}\cdot\ln\left(\frac{1-t}{2}\right)$\\[0.5em]
$n=4$, $L=4$&$\frac{577-5549t-6406t^2+24886t^3+8069t^4-21929t^5}{1056(1-t)}+\frac{5(1-14t^2+21t^4)}{8}\cdot\ln\left(\frac{1-t}{2}\right)$\\[0.5em]
$n=5$, $L=1$&$\frac{(\pi-\vartheta)\cdot(3-12t^2+8t^4)}{8(1-t^2)^{3/2}}+\frac{t(13-16t^2)}{24(1-t^2)}$\\[0.5em]
$n=5$, $L=2$&$\frac{(\pi-\vartheta)\cdot t(15-40t^2+24t^4)}{8(1-t^2)^{3/2}}-\frac{3-23t^2+22t^4}{16(1-t^2)}$\\[0.5em]
$n=5$, $L=3$&$\frac{(\pi-\vartheta)\cdot(-5+60t^2-120t^4+64t^6)}{8(1-t^2)^{3/2}}-\frac{t(37-144t^2+112t^4)}{40(1-t^2)}$\\[0.5em]
$n=5$, $L=4$&$\frac{(\pi-\vartheta)\cdot t(-35+210t^2-336t^4+160t^6)}{8(1-t^2)^{3/2}}+\frac{9-159t^2+416t^4-272t^6}{48(1-t^2)}$\\[0.5em]
$n=6$, $L=1$&$\frac{56+64t-359t^2+232t^3}{105(1-t)^2}+t\cdot\ln\left(\frac{1-t}{2}\right)$\\[0.5em]
$n=6$, $L=2$&$\frac{-103+692t+6t^2-1844t^3+1237t^4}{180(1-t)^2}-2(1-7t^2)\cdot\ln\left(\frac{1-t}{2}\right)$\\[0.5em]
$n=6$, $L=3$&$\frac{-704-1519t+11288t^2-3342t^3-18464t^4+12697t^5}{660(1-t)^2}-\frac{7t(1-3t^2)}{2}\cdot\ln\left(\frac{1-t}{2}\right)$\\[0.5em]
$n=6$, $L=4$&$\frac{5477-58742t-30293t^2+384684t^3-166405t^4-450454t^5+315317t^6}{6240(1-t)^2}
   +\frac{7(1-18t^2+33t^4)}{8}\cdot\ln\left(\frac{1-t}{2}\right)$\\[0.5em]
$n=7$, $L=1$&$\frac{(\pi-\vartheta)\cdot(5-30t^2+40t^4-16t^6)}{16(1-t^2)^{5/2}}+\frac{t(35-84t^2+46t^4)}{48(1-t^2)^2}$\\[0.5em]
$n=7$, $L=2$&$\frac{(\pi-\vartheta)\cdot(35-140t^2+168t^4-64t^6)}{16(1-t^2)^{5/2}}-\frac{62-695t^2+1304t^4-656t^6}{240(1-t^2)^2}$\\[0.5em]
$n=7$, $L=3$&$\frac{(\pi-\vartheta)\cdot(-35+560t^2-1680t^4+1792t^6-640t^8)}{48(1-t^2)^{5/2}}-\frac{t(255-1462t^2+2240t^4-1024t^6)}{144(1-t^2)^2}$\\[0.5em]
$n=7$, $L=4$&$\frac{(\pi-\vartheta)\cdot t(-105+840t^2-2016t^4+1920t^6-640t^8)}{16(1-t^2)^{5/2}}
   +\frac{122-2831t^2+10960t^4-14160t^6+5888t^8}{336(1-t^2)^2}$\\[0.5em]
$n=8$, $L=1$&$\frac{144+131t-1518t^2+2013t^3-776t^4}{315(1-t)^3}+t\cdot\ln\left(\frac{1-t}{2}\right)$\\[0.5em]
$n=8$, $L=2$&$\frac{-2927+23367t-14646t^2-75134t^3+114273t^4-45021t^5}{4620(1-t)^3}-\frac{1-9t^2}{2}\cdot\ln\left(\frac{1-t}{2}\right)$\\[0.5em]
$n=8$, $L=3$&$\frac{-6656-13551t+155859t^2-155858t^3-250170t^4+450453t^5-180181t^6}{5460(1-t)^3}-\frac{3t(3-11t^2)}{2}\cdot\ln\left(\frac{1-t}{2}\right)$\\[0.5em]
$n=8$, $L=4$&$\frac{4173-49699t+5793t^2+402945t^3-485545t^4-379929t^5+843387t^6-341189t^7}{3360(1-t)^3}
   +\frac{3(3-66t^2-143t^4)}{8}\cdot\ln\left(\frac{1-t}{2}\right)$\\\hline
\end{tabular}\end{table}

\addtocounter{section}{1}

\section*{Appendix}

In this section, we collect several hints for computing integrals of irrational functions that arise when one wants to derive a formula for the kernel~$\mathcal K$ in an even-dimensional space.

\begin{lem} If $\lambda$ is a half-integer, $\lambda\in\mathbb N/2\setminus\mathbb N$, then the following holds:
\begin{align*}
&\int\left[\frac{1-r^2}{r(1-2tr+r^2)^{\lambda+1}}-\frac{1}{r}\right]dr\\
&=\frac{1}{\lambda(\mathbf t+\mathbf r^2)^\lambda}+\frac{1}{2}\sum_{j=1}^{\lambda-\frac{1}{2}}\frac{1}{(\lambda-j)(\mathbf t+\mathbf r^2)^{\lambda-j}}
   +\sum_{j=1}^{\lambda-\frac{1}{2}}\frac{t\,Q^\lambda_{j-1}(\mathbf t)\,\mathbf r}{\mathbf t^j(\mathbf t+\mathbf r^2)^{\lambda-j}}
   -\ln\left(\mathbf t-t\mathbf r+\sqrt{\mathbf t+\mathbf r^2}\right)+C,
\end{align*}
where $\mathbf r=r-t$ and $\mathbf t=1-t^2$, and polynomials $Q^\lambda_j$, $j=0,1,\dots,\lambda-\frac32$, are given recursively by
\begin{align*}
1=&\,2(\lambda-1)Q^\lambda_0(\mathbf t),\\
0=&\,2(\lambda-2)Q^\lambda_1(\mathbf t)-\left[(2\lambda-3)+2(\lambda-1)\mathbf t\right]Q^\lambda_0(\mathbf t),\\
0=&\,2(\lambda-j-1)Q^\lambda_j(\mathbf t)-\left[(2\lambda-2j-1)+2(\lambda-j)\mathbf t\right]Q^\lambda_{j-1}(\mathbf t)\\
   &+(2\lambda-2j+1)\mathbf tQ^\lambda_{j-2}(\mathbf t),\hspace{10em}j=2,\dots,\lambda-\tfrac{3}{2}.
\end{align*}\\
\end{lem}

\begin{lem}Suppose $\mathbf t\ne0$ and consider the integral
$$
I_{k,J+\frac12}:=\int\frac{\mathbf R^k}{(\mathbf t+\mathbf R^2)^{J+\frac12}}\,d\mathbf R.
$$
If $k=2\kappa+1$, then
$$
I_{k,J+\frac12}=\sum_{\iota=0}^\kappa\binom{\kappa}{\iota}
   \frac{(-1)^{\kappa-\iota+1}}{2(J-\iota)-1}\cdot\frac{\mathbf t^{\kappa-\iota}}{(\mathbf t+\mathbf R^2)^{J-\iota-\frac12}}+C.
$$
If $k=2\kappa$ and $\kappa<J$, then
$$
I_{k,J+\frac12}=\mathbf t^{\kappa-J}\sum_{\iota=0}^{J-\kappa-1}\binom{J-\kappa-1}{\iota}\frac{(-1)^{J-\kappa-\iota-1}}{2(J-\iota)-1}
   \cdot\left(\frac{\mathbf R^2}{\mathbf t+\mathbf R^2}\right)^{J-\iota-\frac12}+C.
$$
If $k=2\kappa$ and $\kappa\geq J$, then
$$
I_{k,J+\frac12}=\frac{\mathbf R}{(\mathbf t+\mathbf R^2)^{J-\frac12}}\cdot P_{k,J+\frac12}(\mathbf t,\mathbf R^2)
   +a\,\mathbf t^{\kappa-J}\,\ln\left(\mathbf R+\sqrt{\mathbf t+\mathbf R^2}\right)+C,
$$
where
$$
a^{\kappa,J+\frac12}=\frac{(-1)^{\kappa-J}(2\kappa-1)!!}{2^{\kappa-J}\cdot(\kappa-J)!\cdot(2J-1)!!},
$$
$P_{k,J+\frac12}$ is a homogeneous polynomial of degree~$\kappa-1$ given by
$$
P_{k,J+\frac12}(\mathbf t,\mathbf R^2)=\sum_{\iota=0}^{\kappa-1} a_\iota^{\kappa,J+\frac12}\mathbf t^{\kappa-\iota-1}\,\mathbf R^{2\iota}
$$
with coefficients obtained recurrently
\begin{align*}
a_0^{\kappa,J+\frac12}&=-a^{\kappa,J+\frac12},\\
a_1^{\kappa,J+\frac12}&=-\frac{(3J-2)\,a^{\kappa,J+\frac12}}{3},\\
a_\iota^{\kappa,J+\frac12}&=\frac{2(J-\iota)\,a_{\iota-1}^{\kappa,J+\frac12}-a^{\kappa,J+\frac12}\binom{J}{\iota}}{2\iota+1},\qquad\iota=2,3,\dots,\kappa-1,
\end{align*}
and $P_{0,J+\frac12}\equiv0$ (with the conventions $(-1)!!=1$ and $\binom{J}{\iota}=0$ for $\iota>J$).\\
\end{lem}

\begin{lem}\label{lem:integral_LJ12}The integral
$$
\mathcal I_{L,J+\frac12}(t,R):=\int\frac{R^L}{(1-2tR+R^2)^{J+\frac12}}\,dR,
$$
$L,J\in\mathbb N_0$, is equal to
$$
\mathcal I_{L,J+\frac12}=\frac{A_{L,J+\frac12}(t,R)}{(1-2tR+R^2)^{J-\frac{1}{2}}\cdot(1-t^2)^J}+B_{L,J+\frac12}(t)\cdot\ln\left(R-t+\sqrt{1-2tR+R^2}\right)+C,
$$
where the polynomials $A$ and $B$ are given by
\begin{align*}
A_{L,J+\frac12}&(t,R)=\sum_{\kappa=0}^{J-1}\sum_{\iota=0}^{J-\kappa-1}\alpha_{\kappa,\iota}^{L,J+\frac12}\,
   t^{L-2\kappa}\,\mathbf t^\kappa\,\mathbf R^{2(J-\iota-\frac12)}(\mathbf t+\mathbf R^2)^\iota\\
&+\sum_{\kappa=J}^{\left[\frac{L}{2}\right]}\sum_{\iota=0}^{\kappa-1}\beta_{\kappa,\iota}^{L,J+\frac12}\,t^{L-2\kappa}\,\mathbf t^{J+\kappa-\iota-1}\,\mathbf R^{2\iota+1}
   +\sum_{\kappa=0}^{\left[\frac{L-1}{2}\right]}\sum_{\iota=0}^\kappa\gamma_{\kappa,\iota}^{L,J+\frac12}\,t^{L-2\kappa-1}\,\mathbf t^{J+\kappa-\iota}\,(\mathbf t+\mathbf R^2)^\iota\\
B_{L,J+\frac12}&(t)=\sum_{\kappa=J}^{\left[\frac{L}{2}\right]}\mu_\kappa^{L,J+\frac12}\cdot t^{L-2\kappa}\cdot\mathbf t^{\kappa-J},
\end{align*}
with $\mathbf R=R-t$, $\mathbf t=1-t^2$, and
\begin{align*}
\alpha_{\kappa,\iota}^{L,J+\frac12}&=\binom{L}{2\kappa}\binom{J-\kappa-1}{\iota}\frac{(-1)^{J-\kappa-\iota-1}}{2(J-\iota)-1},\\
\beta_{\kappa,\iota}^{L,J+\frac12}&=\binom{L}{2\kappa}a_\iota^{\kappa,J+\frac12},\\
\gamma_{\kappa,\iota}^{L,J+\frac12}&=\binom{L}{2\kappa+1}\binom{\kappa}{\iota}\frac{(-1)^{\kappa-\iota+1}}{2(J-\iota)-1},\\
\mu_\kappa^{L,J+\frac12}&=\binom{L}{2\kappa}a^{\kappa,J+\frac12}.
\end{align*}
\end{lem}

{\bf Proof.} Follows from
\begin{align*}
\mathcal I_{L,J+\frac12}&=\int\frac{(\mathbf R+t)^L}{(\mathbf t+\mathbf R^2)^{J+\frac12}}\,d\mathbf R=\sum_{k=0}^L\binom{L}{k}t^{L-k}I_{k,J+\frac12}\\
&=\sum_{\kappa=0}^{\min\left\{J,\left[\tfrac{L}{2}\right]\right\}-1}\binom{L}{2\kappa}t^{L-2\kappa}I_{2\kappa,J+\frac12}
   +\sum_{\kappa=J}^{\left[\frac{L}{2}\right]}\binom{L}{2\kappa}t^{L-2\kappa}I_{2\kappa,J+\frac12}\\
&+\sum_{\kappa=0}^{\left[\frac{L-1}{2}\right]}\binom{L}{2\kappa+1}t^{L-2\kappa-1}I_{2\kappa+1,J+\frac12}.
\end{align*}
\hfill$\Box$

\begin{lem}
Let $k$ be a positive integer. The integral
$$
\int R^k\ln\left(1-tR+\sqrt{1-2tR+R^2}\right)\,dR
$$
equals
\begin{align*}
&p_k(t,R)\cdot\sqrt{1-2tR+R^2}+q_k(t)\cdot\ln\left(R-t+\sqrt{1-2tR+R^2}\right)\\
&+\frac{R^{k+1}}{k+1}\left[\ln\left(1-tR+\sqrt{1-2tR+R^2}\right)-\frac{1}{k+1}\right]+C,
\end{align*}
where
$$
p_k(t,R)=\sum_{j=0}^{k-1}R^j\cdot\pi_j^k(t)
$$
with $\pi_j^k$ defined recursively by
\begin{align*}
\pi_{k-1}^k(t)&=\frac{1}{k(k+1)},\\
\pi_{k-2}^k(t)&=\frac{2k-1}{k-2}\,t\pi_{k-1}^k(t),\\
\pi_j^k(t)&=\frac{1}{j+1}\left[(2j+3)t\pi_{j+1}^k(t)-(j+2)\pi_{j+2}^k(t)\right],&j=k-3,k-4,\dots1,\\
\pi_0^k(t)&=3t\pi_1^k(t)-2\pi_2^k(t),
\end{align*}
and
$$
q_k(t)=t\pi_0^k(t)-\pi_1^k(t).
$$
\end{lem}


\vspace{3em}
\begin{thebibliography}{99}
\bibitem{AAMASD21} R. Amina, B. Alshahrani, M. Mahmoud, A.-H. Abdel-Aty, K. Shah, and W. Deebanif, \emph{Haar wavelet method for solution of distributed order time-fractional differential equations}, Alexandria Engineering J. (60) 2021, 3295--3303.
\bibitem{AV} J.-P. Antoine and P. Vandergheynst, \emph{Wavelets on the 2-sphere: a group-theoretical approach}, Appl. Comput. Harmon. Anal. 7 (1999), No.~3, 262--291.
\bibitem{AVn} J.-P. Antoine and P. Vandergheynst, \emph{Wavelets on the n-sphere and related manifolds}, J. Math. Phys. 39 (1998), No.~8, 3987--4008.
\bibitem{BH18} S. Balaji and G. Hariharan, \emph{A novel wavelet approximation method for the solution of nonlinear differential equations with variable coefficients arising in astrophysics}, Astrophys Space Sci 363, 16 (2018), https://doi.org/10.1007/s10509-017-3236-3.
\bibitem{sB09} S.~Bernstein, \emph{Spherical singular integrals, monogenic kernels and wavelets on the three--dimensional sphere}, Adv. Appl. Clifford Algebr. 19 (2009), No.~2, 173--189.
\bibitem{BZ09} J.P. Boyd and C. Zhou, \emph{Three ways to solve the Poisson equation on a sphere with Gaussian forcing}, J. Comput. Phys. 228 (2009), 4701--4713.
\bibitem{oC03} O.~Christensen, \emph{An introduction to frames and {R}iesz bases}, Birkh\"auser, Boston 2003.
\bibitem{CJ99} O.~Christensen and T.K.~Jensen, \emph{An introduction to the theory of bases, frames, and wavelets},
http://citeseerx.ist.psu.edu/viewdoc/summary?doi=10.1.1.17.241, 1999.
\bibitem{DX13} F.~Dai, Y.Xu, \emph{Approximation Theory and Harmonic Analysis on Spheres and Balls}, Springer, New York 2013.
\bibitem{EBCK09} S. Ebert, S. Bernstein, P. Cerejeiras, and U. K\'ahler, \emph{Nonzonal wavelets on~$\mathcal S^N$}, 18$^\text{th}$ International Conference on the Application of Computer Science and Mathematics in Architecture and Civil Engineering, Weimar 2009.
\bibitem{mF99} M.W. Frazier, \emph{An introduction to wavelets through linear algebra}, Springer, New York, 1999.
\bibitem{FGS-book} W. Freeden, T. Gervens, and M. Schreiner, \emph{Constructive approximation on the sphere. With applications to geomathematics}, Clarendon Press, New York 1998.
\bibitem{FW-C} W.~Freeden and U.~Windheuser, \emph{Combined spherical harmonic and wavelet expansion -- a future concept in {E}arth's gravitational determination}, Appl. Comput. Harmon. Anal. 4 (1997), No. 1, 1--37.
\bibitem{FW} W.~Freeden and U.~Windheuser, \emph{Spherical wavelet transform and its discretization}, Adv. Comput. Math. 5 (1996), No.~1, 51--94.
\bibitem{cjG14} C.J. Gittelson, \emph{Adaptive wavelet methods for elliptic partial differential equations with random operators},  Numer. Math. 126 (2014), no. 3, 471--513.
%\bibitem{GR} I.S. Gradshteyn and I.M.~Ryzhik, \emph{Table of integrals, series, and products}, Elsevier/Academic Press, Amsterdam, 2007.
\bibitem{GS04} R. Grebenitcharsky and M.G. Sideris M.G., \emph{Application of spherical pseudo-differential operators and spherical wavelets for numerical solutions of the fixed altimetry-gravimetry boundary value problem}. In: F. Sansò (eds) \emph{V Hotine-Marussi Symposium on Mathematical Geodesy. International Association of Geodesy Symposia}, vol 127. Springer, Berlin, Heidelberg. https://doi.org/10.1007/978-3-662-10735-5\_31.
\bibitem{kG01} K.~Gr\"ochenig, \emph{Foundations of Time-Frequency Analysis}, Applied and Numerical Harmonic Analysis, Birkh\"auser, Boston, 2001.
\bibitem{GR14} A.K. Gupta and S.S. Ray, S. Saha, \emph{Wavelet methods for solving fractional order differential equations}, Math. Probl. Eng. 2014, Art. ID 140453, 11 pp.
\bibitem{HMV04} M.A. Hajji, S. Melkonian, and R. Vaillancourt, \emph{Two-dimensional wavelet bases for partial differential operators and applications}, Advances in pseudo-differential operators, 219--233, Oper. Theory Adv. Appl., 155, Birkhäuser, Basel, 2004.
\bibitem{HMW19} B. Han, M. Michelle, and Y.S. Wong, \emph{Wavelet-based Methods for Numerical Solutions of Differential Equations}, arXiv:1909.12192.
\bibitem{HK10} G. Hariharan and K. Kannan, \emph{Haar wavelet method for solving some nonlinear parabolic equations}, J. Math. Chem. 48 (2010), no. 4, 1044--1061.
\bibitem{IIN19CWTforC} I. Iglewska-Nowak, \emph{A continuous spherical wavelet transform for~$\mathcal C(\mathbb R^n)$},  Appl. Comput. Harmon. Anal. 47 (2019), no. 3, 1033--1039.
\bibitem{IIN15CWT} I. Iglewska-Nowak, \emph{Continuous wavelet transforms on $n$-dimensional spheres}, Appl. Comput. Harmon. Anal. 39 (2015), no. 2, 248--276.
\bibitem{IIN18DW}  I.~Iglewska--Nowak, \emph{Directional wavelets on $n$-dimensional spheres},  Appl. Comput. Harmon. Anal. 44 (2018), no. 2, 201--229.
\bibitem{IIN17FDW} I.~Iglewska--Nowak, \emph{Frames of directional wavelets on $n$--dimensional spheres}, Appl. Comput. Harmon. Anal. 43 (2017), 148--161.
\bibitem{IIN15PW} I. Iglewska-Nowak, \emph{Poisson wavelets on $n$-dimensional spheres}, J. Fourier Anal. Appl. 21 (2015), no. 1, 206--227.
\bibitem{sJ92} S. Jaffard, \emph{Wavelets and analysis of partial differential equations}, in: J.S. Byrnes, J.L. Byrnes, K.A. Hargreaves, and K. Berry, \emph{Probabilistic and stochastic methods in analysis, with applications}, Springer Science+Bussiness Media Dodrecht, 1992.
\bibitem{gK94} G.~Kaiser, \emph{A friendly guide to wavelets}, Birkh\"auser, Boston, 1994.
\bibitem{LH14} \"U. Lepik and H. Hein, \emph{Haar wavelets. With applications}, Mathematical Engineering. Springer, Cham, 2014.
\bibitem{MLZJ05} S.L. Mei, Q.S. Lu, S.W. Zhang, and L. Jin, \emph{Adaptive interval wavelet precise integration method for partial differential equations}, Appl. Math. Mech. 26 (2005), no. 3, 364--371.
\bibitem{vP90} V. Perrier, \emph{Towards a Method for Solving Partial Differential Equations Using Wavelet Bases}, in: J.-M. Combes, A. Grossman, and P. Tchamitchian (Eds.) \emph{Wavelets}, Springer, Berlin, Heidelberg, 1989.
\bibitem{RBAIS20} M. Rehman, D. Baleanu, J. Alzabut, M. Ismail, and U. Saeed, \emph{Green–Haar wavelets method for generalized fractional differential equations}, Advances in Difference Equations (2020), 2020:515.
\bibitem{SYYS98} J. Sun, X. Yi, B. Ye, and Y. Shen, \emph{Wavelet-Galerkin Solutions for Differential Equations}, Wuhan University Journal of Natural Scmnces 3 (1998), no. 4 1998, 403--406.
\bibitem{Shimakura} N. Shimakura, \emph{Partial differential operators of elliptic type}, Translations of Mathematical Monographs, Vol. 99, Amer. Math. Soc., Providence, Rhode Island, 1992.
\bibitem{Sz75} G.~Szeg\"o, \emph{Orthogonal polynomials}, Fourth Edition, AMS Coll. Publ., Providence, Rhode Island, 1975.
\bibitem{rSz06} R. Szmytkowski, \emph{Closed form of the generalized Green's function for the Helmholtz operator on the two-dimensional unit sphere}, J. Math. Phys. 47 (2006), no. 6, 063506, 11 pp.
\bibitem{rSz07} R. Szmytkowski, \emph{Closed forms of the Green's function and the generalized Green's function for the Helmholtz operator on the N-dimensional unit sphere}, J. Phys. A 40 (2007), no. 5, 995--1009.
\bibitem{kU09} K. Urban, \emph{Wavelet methods for elliptic partial differential equations}, Numerical Mathematics and Scientific Computation. Oxford University Press, Oxford, 2009.
\bibitem{Vilenkin} N. Ja. Vilenkin, \emph{Special functions and the theory of group representations}, in Translations of Mathematical Monographs, Vol. 22, American Mathematical Society, Providence, R. I., 1968.
\bibitem{WMM14} L. Wang, Y. Ma, and Z. Meng, \emph{Haar wavelet method for solving fractional partial differential equations numerically}, Appl. Math. Comput. 227 (2014), 66--76.
\bibitem{ZXZ16} S. Zhi, Y. Xu, and J. Zhao, Jun-ping, \emph{Haar wavelets method for solving Poisson equations with jump conditions in irregular domain}, Adv. Comput. Math. 42 (2016), no. 4, 995--1012.
\end{thebibliography}
\end{document}